\documentclass[11pt]{article}

\usepackage{fullpage}
\usepackage{xcolor}
\usepackage{natbib}

\usepackage{amsmath,amssymb,amsthm}
\newtheorem{definition}{Definition}

\definecolor{DarkGreen}{rgb}{0.1,0.5,0.1}

\title{Integrating Differential Privacy and Contextual Integrity}

\author{Sebastian Benthall\thanks{New York University School of Law. Email: \texttt{spb413@nyu.edu}} \and Rachel Cummings\thanks{Columbia University. Email \texttt{rac2239@columbia.edu}} }
\date{}

\begin{document}

\maketitle

\begin{abstract}
In this work, we propose the first framework for integrating Differential Privacy (DP) and Contextual Integrity (CI). DP is a property of an algorithm that injects statistical noise to obscure information about individuals represented within a database. CI defines privacy as information flow that is appropriate to social context. Analyzed together, these paradigms outline two dimensions on which to analyze privacy of information flows: descriptive and normative properties. We show that our new integrated framework provides benefits to both CI and DP that cannot be attained when each definition is considered in isolation: it enables contextually-guided tuning of the $\epsilon$ parameter in DP, and it enables CI to be applied to a broader set of information flows occurring in real-world systems, such as those involving PETs and machine learning. We conclude with a case study based on the use of DP in the U.S. Census Bureau.

\end{abstract}

\section{Introduction}\label{s.intro}

Contextual Integrity (CI) \citep{nissenbaum2004privacy} is a privacy paradigm that emphasizes contextual, normative properties of information flows, and defines privacy as maintaining information flows that are appropriate to social context. Differential Privacy (DP) \citep{DMNS06}, on the other hand, is a property of an algorithm that injects statistical noise to obscure information about individuals' data stored in a database. This privacy notion emphasizes descriptive algorithmic and mathematical properties of information flows without any contextual considerations. Although these privacy frameworks both relate to privacy of information flows, they have previously only been considered separately. Analyzed together, these frameworks outline two dimensions of criteria on which to analyze systems for privacy: descriptive and normative properties; and information-flow-specific and contextual properties.

In this paper, we explore the integration of CI and DP paradigms to enable reasoning about applications of DP -- and other privacy enhancing technologies (PETs) -- based on the context of deployment. We propose a new framework that augments contextual integrity to include a new \emph{transmission property} feature that captures algorithmic applications of PETs, such as DP, to information flows. 

We argue that this augmented framework provides benefits to both research communities that cannot be attained when each definition is considered in isolation. Specifically, it enables context-specific parameter tuning of the privacy parameter $\epsilon$ in differential privacy, which must be tuned to balance the trade-off between accuracy and privacy of analysis. While it is widely agreed among researchers that this parameter should vary based on the context, there is very little guidance for how to determine appropriate values of $\epsilon$ \citep{dwork2019differential}.

In the reverse direction, our integrated framework expands the scope of information flows that can be considered under CI to better match information flows occurring in real-world systems. Whereas the original CI definition only allowed unitary data flows (i.e., share complete data or not) about a single data subject, our new framework allows CI to be applied to analysis of databases containing data from multiple individuals and to incorporate uncertainty either from PETs or from statistical sources such as sampling error. Explicitly, this expands the applicability of CI in two important dimensions: to include information flows about \emph{multiple people} and about \emph{functions of the data} rather than just directly transmitting the data. Both of these dimensions are standard in both modern machine learning practices and in PETs use (e.g., DP analysis of a database). Although we focus our attentions on DP in this paper, many other PETs and machine learning techniques could be used as transmission properties in our new framework.

We conclude with an application of our framework to a case study based on the U.S. Census. We show that the combined language of CI and DP enables better articulation of the (legally grounded) norms surrounding privacy and accuracy of Decennial Census data. The addition of transmission properties into the Census context is \emph{necessary} to identify key distinctions between swapping and DP that determine appropriateness of use of each PET in this context.

\section{Background on Contextual Integrity and Differential Privacy}\label{s.background}

\subsection{Contextual Integrity}\label{s.ci}

\citet{nissenbaum2009privacy} developed Contextual Integrity (CI)
as a theory of socially meaningful privacy that can bridge
between legal and political theory, user interaction design,
and computer science.
According to CI, privacy is appropriate information flow, where appropriateness is defined in terms of contextually grounded information norms.

When it was first proposed, CI was a novel and significant intervention in legal and other scholarship about privacy for at least two reasons.
One reason was that it stipulated a notion of privacy that was distinct from individual control over data.
To this day, CI-oriented scholars remain skeptical of individual control as a solution to concerns about data privacy.
Another reason is that CI's concept of `privacy' perhaps uniquely comprehends that society depends on both positive and negative rules about personal information flow.
While it is a violation of privacy if doctors betray their patient's confidentiality, CI also considers it to be a
violation of privacy if patients were somehow unable to communicate about their health with their doctors.
The constellation of norms in a social context are aimed at enabling appropriate flow and disabling inappropriate flow.
In practice, these two goals are often in tension.

While \citet{nissenbaum2009privacy} did not include formal definitions of CI, for the purpose of this article we develop a formalism for CI that reflects its key concepts and constraints.
A foundational concept in Contextual Integrity is the \emph{context}, which is a normatively understood field of human behavior.

\begin{definition}[Context]\label{def.context}
A context consists of: a set of contextual purposes, a set of agent roles, a set of information attributes, a set of information norms, and a set of transmission principles.
\end{definition}

% \begin{definition}[Context]\label{def.context}
% A context is a defined as a tuple $(P, R, A, N,E)$ where:
% \begin{itemize}
%     \item $P$ are a set of contextual purposes
%     \item $R$ is a set of agent roles
% %    \item $U = \{ U_r | r \in R \}$, the agents \emphs{ends}, are utility functions for each role in $R$
%     \item $A$ is a set of information attributes
%     \item $N$ is a set of information norms
%     \item $E$ is a set of transmission principles
% \end{itemize}
% \end{definition}

Contexts are defined in terms of their purposes in society.
The \emph{contextual purpose} is, at a high level, an explanation for why the context exists in society. The \emph{agent roles} are the actors who operate in the context and engage in tasks in support of the purpose, and the \emph{information attributes} are the kinds of relevant personal information that flow within the context. It is important to emphasize that any context will have multiple purposes, agent roles, and information attributes.

As a concrete example, in the context of health care, the purposes may be to preserve the health of people, to identify novel disease treatments, and to train new medical professionals. Agent roles include doctors, nurses, patients, insurance carriers, medical researchers, and more. Information attributes in this context include all aspects of patient health records such as the results of medical tests and treatment notes, but also includes accompanying contextually-relevant information such as patient demographics, payment codes, insurance coverage levels, and billing information.

\emph{Information norms} specify the set of appropriate information flows within a context. These norms are only defined in terms of the properties of the context in which they adhere. \emph{Transmission principles} are normative restrictions on information flows that fit the description of the other four parameters; these will be the subject of much discussion in this work.

\begin{definition}[Information norm]\label{def.inorm}
Within a context, an information norm is a tuple $(s, r, u, a, e)$:
\begin{itemize}
    \item $s$ is the sender of the information
    \item $r$ is the receiver of the information
    \item $u$ is the subject of the information
    \item $a$ is the information attribute
    \item $t$ is a transmission principle
\end{itemize}
\end{definition}

The sender $s$, receiver $r$, and subject $u$ are all agent roles in the context. An information norm $(s, r, u, a, t)$ describes information $a$ about subject $u$ flowing from sender $s$ to receiver $r$ under the conditions specified by transmission principle $t$. For example, an information norm in the heathcare context would be: A \underline{patient}'s ($u$) \underline{medical record} ($a$) may be sent from their \underline{primary care physician} ($s$) to a \underline{specialist} ($r$) when \underline{the patient requests a referral} ($t$) to that specialist.

Lastly, in CI, contextual norms are adapted to the society in which they
take part.
In particular, contextual norms support the specific purpose
of their context (e.g., a healthy population) by balancing the ends of individual parties (e.g., doctors, patients, insurers) acting within that context.
They may also be calibrated or inflected by societal values (such as ``freedom'') more broadly, and thus may differ across time and cultures.

\subsection{Differential Privacy}

Differential privacy is a parameterized notion of algorithmic privacy for databases. Informally, it bounds the impact of any one data entry on the result of an algorithmic analysis of the database. Formally, it guarantees that changing any single database element will result in a bounded change in the distribution over results of an analysis of the database.

\begin{definition}[Differential privacy \citep{DMNS06}]\label{def.dp}
An algorithm $\mathcal{M}: \mathcal{X}^n \to \mathcal{R}$ that maps from databases containing $n$ entries, to an arbitrary output range is $(\epsilon,\delta)$-differentially private if for all pairs of databases $X,X'\in \mathcal{X}^n$ that differ in a single element, and for all possible results $\mathcal{S} \subseteq \mathcal{R}$ that may be produced by $\mathcal{M}$, 
\[ \Pr[\mathcal{M}(X) \in \mathcal{S}] \leq e^{\epsilon} \Pr[\mathcal{M}(X) \in \mathcal{S}] + \delta.\]
\end{definition}

Differential privacy guarantees are governed by two parameters: $\epsilon \in [0,\infty)$ and $\delta\in [0,1)$, where smaller parameter values correspond to stronger privacy guarantees. 
Early theoretical work on differential privacy suggested $\epsilon$ as a small constant such as 0.1 or 0.01, or a value diminishing in the size of the database, such as $1/\sqrt{n}$. More recent practitioners have preferred larger $\epsilon$ values in the range of 1 to 8 due to the inherent privacy-accuracy tradeoff, although Apple received significant backlash when it was suspected of using $\epsilon=42$ \citep{TKBWW17}.
The $\delta$ parameter is commonly chosen to be 0, although allowing a strictly positive $\delta$ can improve accuracy in many settings. When it is positive, it is preferred to be extremely small, such as $10^{-6}$ or $\exp(-n)$.

DP is achieved algorithmically by adding random noise at some part of the computation. For example, to compute the average of values in the database in a differentially private manner, one could first compute the true average of the values, and then perturb the result by adding a random noise term. Commonly used noise distributions include Gaussian noise with mean zero and Laplace noise (i.e., a two-sided exponential distribution with mean zero), although these are far from the only possibilities. The magnitude of the noise should depend on $\epsilon$, where smaller $\epsilon$ (i.e., stronger privacy) corresponds to more noise being added.

Adding noise inherently comes at a cost of accuracy of the results, since the algorithm is not allowed to output the true answer directly. The relationship between $\epsilon$, the noise parameters, and accuracy of the results, immediately gives a quantifiable privacy-accuracy trade-off, where stronger privacy guarantees necessarily provide lower accuracy. For most common statistical and data science problems, there exists a DP algorithm for privately solving it, and a known mathematical expression relating $\epsilon$ to accuracy of the results.
Much of the computer science literature on DP is devoted to finding creative ways to add sufficient noise in the algorithms, while still maintaining accurate statistical results, thus improving the limits of the privacy-accuracy trade-off.

Differential privacy also has variant models, which capture different trust models of the data holder. The \emph{central model} (presented above in Definition \ref{def.dp}) assumes there is a trusted data curator who will collect and hold the raw data of all users, and promises to publish information about the database using only differentially private algorithms. The \emph{local model} \citep{KLNRS11} assumes that the curator is untrusted, and users do not wish to send their raw data directly to the curator. Instead, they add noise to their own data locally before sending this. This model can be interpreted formally as Definition \ref{def.dp} with $n=1$. The \emph{shuffle model} \citep{BEM+17} can be viewed an intermediate model, where users apply DP locally to their own data, and a cryptographically secure ``shuffler'' permutes all users' data. This shuffling step allows for less noise to be added while still maintaining the same accuracy guarantee.

\section{Limitations of Single Definitions}\label{s.limit}

Although both CI and DP provide considerable benefits as methods for providing privacy in real-world settings, they both have limitations when considered in isolation. At a high-level, CI can be applied broadly to all contexts, but does not provide algorithmically actionable privacy interventions, and its formalization of data flows do not necessarily match modern uses of data in machine learning systems. DP, on the other hand, provides concrete algorithmic implementations, but leaves unspecified free parameters that must be tuned based on the context. Both of these limitations are explored in more detail in this section.

\subsection{Limitations of CI}

While CI provides a broad philosophical framework for determining appropriateness of information flows in new and existing contexts, it has some shortcomings that limit its application in modern data use scenarios.

\textbf{1. CI considers a singular data subject with unitary data flow.} The model of CI considers an information flow with a singular, stable information attribute $a$ and data subject $u$. While this \emph{conduit metaphor} \citep{reddy1979conduit} of information flows may be suitable for mapping onto social settings -- e.g., a doctor ($r$) observes a patient's ($u$) medical record ($a$) -- it fails to capture modern data sharing scenarios encountered with machine learning. For example, when the data object transferred is a model trained on a database of information collected from multiple individuals, who is the \emph{data subject} and what is the \emph{information attribute} that is being conveyed? How can uncertainty about the information that is shared -- either due to sampling error in the data collection or due to privacy protections such as DP -- be incorporated into the \emph{information attribute}? 

Prior work within the CI literature has also offered a number of related internal critiques. \citet{benthall2019situated} highlighted that information flows may be situated within a larger network of other information flows, such that these flows may themselves be a part of a larger context. The notion of a singular data subject is challenged by notions of relational \citep{viljoen2021relational} and group \citep{taylor2016group} privacy, which highlight how data about one individual can support inferences about others who share relevant characteristics, especially when machine learning and statistical techniques are employed.
Indeed, it is these latter cases which are among the motivations for privacy enhancing technologies such as differential privacy.

\textbf{2. No general framework for applying CI.} Contextual integrity does not come with algorithmic approaches for achieving it that can be easily applied in new contexts. Rather, when the framework of CI is applied to a novel context, considerable work must be done by CI experts to understand the social norms and expectations of information flows in that context. Even when the social and cultural norms of a context are fully known, CI does not have algorithmic techniques that can be applied to \emph{solve}, \emph{reduce}, or \emph{prevent} privacy challenges, but rather to \emph{identify} whether the information flows constitute privacy violations.

\subsection{Limitations of DP}

On the other hand, DP provides a suite of algorithmic tools for ensuring parameterized privacy guarantees in any context, but it also has shortcomings that limit its usefulness in real-world applications.

\textbf{1. DP offers no guidance on tuning parameters.} The privacy guarantees of DP are controlled by the parameter $\epsilon$, which can range from 0 (corresponding to perfect privacy) to $\infty$ (corresponding to no privacy). The value of $\epsilon$ also has implications for the the accuracy of analysis. Most theoretical papers in the DP literature present the accuracy guarantees as a function of $\epsilon$, and then posit that the value of $\epsilon$ should be chosen to balance the privacy and accuracy needs of the context, but does not provide any further guidance. The problem of practical and context-specific guidance for choosing $\epsilon$ is one of the largest barriers to widespread deployment of DP today \citep{dwork2019differential,nanayakkara2023chances,cummmings2023centering}.

\textbf{2. DP is agnostic to social norms and context.} While DP focuses on mathematical precision of the algorithm's properties, it is purposefully agnostic to the context or societal norms within which it is deployed. The DP community has historically been largely silent on how this definition integrates with societal concerns. On the one hand, this can be seen as a feature that enables DP to serve as a general-purpose privacy tool; on the other hand, this causes the field of DP to lack guidance for how the suite of algorithmic tools should be applied, tailored, and adapted in new contexts. These questions include the choice of privacy parameters as described above, but it also includes questions of whether DP is appropriate for use in a given context, the most appropriate DP algorithm for the task, how to allocate a given privacy budget across multiple analyses of the same database, and whether the central or local model is more appropriate for the context. These are ultimately normative questions surrounding the use of a descriptive toolkit.

\section{How to bridge CI and DP}\label{s.bridge}

Our main contribution is an integrated formalism of CI and DP that addresses the challenges identified in Section \ref{s.limit} by building upon the framework of CI. In short, while CI captures the \emph{normative} aspects of privacy in an information system, it does not allow for language about \emph{descriptive} properties of the algorithmic system. On the other hand, the DP definition is entirely \emph{descriptive} without any discussion of norms, but leaves unanswered the surrounding \emph{normative} questions around details of appropriate deployments in varying contexts. We argue that effective privacy-enhancing systems require integration of both normative and descriptive aspects of privacy; appropriate information flows should be tailored to both the descriptive and normative aspects of a context. 

In Section \ref{s.trans}, we introduce an additional CI parameter called \emph{transmission properties}, which are the descriptive technical properties of an information flow, and can capture algorithmic privacy properties like the use of differential privacy. This expansion of scope of the qualitative information used to analyze privacy of an information systems has benefits for both the CI and DP communities: it enables the CI framework to extend to modern uses of data, where guarantees are probabilistic and analysis occurs over a larger dataset, rather than a unitary data flow. Perhaps more importantly, this formalism also enables the language of context and social norms to be applied to deployments of differential privacy, and can be used to tune DP parameters (such as $\epsilon$) to both ensure an appropriate information flow in a specific context, and to facilitate optimal parameter tuning within the space of appropriate choices. We emphasize that while we focus on the integration of DP and CI in this work, the notion of transmission principles can also be applied to other algorithmic PETs, such as encrypted communication, secure multi-party computation, federated learning, data minimization, and more.

We acknowledge one major challenge is that normative social expectations have not fully coalesced  around the use of DP. This is due to several factors, including differences between people's internal mental models of privacy \citep{camp2009mental} and the nature of the privacy guarantees offered by DP, and the fact that many people do not understand how DP works \citep{cummings2021need,sarathy2023dont}, and thus are unable to consider it within their normative expectations of information flow. While fully addressing these challenges is beyond the scope of the current work, it has been the subject of much recent study \citep{nanayakkara2023chances,smart2023model,DCHL23}, and we hope that this framework can help guide the discussion of appropriate choices of DP parameters across various contexts. As norms surrounding the use of differential privacy continue to evolve and become better understood, our integrated framework can be used to determine appropriateness of deployment parameters in existing and future contexts.

\subsection{Integration of CI and DP using Transmission Properties}\label{s.trans}

Formally, we propose to bridge the notions of CI and DP by adding an additional \emph{transmission property} parameter to the definitions of a Context (Definition \ref{def.context}) and an Information Norm (Definition \ref{def.inorm}). The transmission properties include any algorithmic transformations or operations that may be performed on the information that may obscure or modify its content in order to provide additional privacy protections. This includes the full suite of PETs (and their implicit parameters) that may be applied to an information flow.

\begin{definition}[Context, Augmented]\label{def.contextaug}
A context consists of: a set of contextual purposes, a set of agent roles, a set of information attributes, a set of information norms, a set of transmission principles, \textbf{and a set of transmission properties}.
\end{definition}

\begin{definition}[Information norm, Augmented]\label{def.inormaug}
Within a context, an information norm is a tuple $(s, r, u, a, e, p)$:
\begin{itemize}
    \item $s$ is the sender of the information
    \item $r$ is the receiver of the information
    \item $u$ is the subject of the information
    \item $a$ is the information attribute
    \item $t$ is a transmission principle
    \item \textbf{$p$ is a transmission property}
\end{itemize}
\end{definition}

Transmission properties cover \emph{descriptive} aspects of a flow, such as the algorithmic mechanics of generating and executing the flow. Examples of transmission properties include: ``with the addition of Gaussian noise of mean 0 and variance $1$,'' ``in the local model of DP with $\epsilon=1/2$ using the Randomized Response Mechanism,'' ``in the shuffle model of DP with $\epsilon=2$ and a secure shuffler,'' or ``with no PETs.''\footnote{Again we emphasize that although we focus on differential privacy as our primary PET of interest in this work, the notion of transmission properties naturally extends to describe the algorithmic and mathematical properties of other PETs and machine learning tools.} We contrast this with the existing CI parameter of transmission \emph{principles }, which emphasize \emph{normative} aspects of a flow, such as ``with subject consent'' or ``under a reciprocity agreement between sender and receiver.''

\begin{table*}[tbh]
\centering
\begin{tabular}{|lll|}
\hline
Transmission Properties & \hspace{3mm} &Transmission Principles \\
\emph{(Descriptive)} & &\emph{(Normative)}  \\
\hline
Flow with no PET & & With subject consent  \\
With Gaussian noise $\mathcal{N}(0,1)$ added & &  Under reciprocity agreement \\
Within an 95\% confidence interval & & Constitutes disclosure \\
Using public-key encryption & & With a warrant   \\
With secure multi-party computation & & As mandated by law \\
\hline
\end{tabular}
\caption{Examples of Transmission Properties and Transmission Principles.}
\label{tab:integrated}
\end{table*}

Table \ref{tab:integrated} provides further examples of of transmission properties involving DP and other PETs. It also provides more examples of transmission principles, to highlight the contrast between normative and descriptive aspects of information flows.

The distinction between normative and descriptive properties is well-founded in the CI literature. \citet{benthall2017contextual} characterize two divergent ways of conceptualizing the notion of ``context''
in the computer science literature: the descriptive \emph{situation}
of the system, including its users and their location; and
the normative social \emph{sphere} of social expectations
in which norms are embedded. The situation of a system includes the size of the populations involved, the distribution of any heterogeneous features, and any specific applicable threat model; on the other hand, examples of (normative) spheres include ``healthcare,'' ``financial systems,'' and ``education.'' Our approach to consider both the normative and descriptive elements of an information flow can be combined with the framework of \citet{benthall2017contextual} to simultaneously take into account both the normative and descriptive aspects of both the \emph{context} of a social environment and technology, as well as the \emph{flows} of information within it.

\subsection{Benefits of Integrated Approach}

The fields of CI and DP have primarily been studied in parallel, and an integrated approach that bridges these two definitions can provide benefits to both academic communities. We highlight some of these benefits here.

\textbf{Guidance for tuning DP parameters.} The mathematical privacy guarantees in the definition of DP are entirely agnostic to normative context of a data flow. Although this has been trumpeted as a strength of DP -- that practitioners can deploy these tools \emph{without} having to reason about context -- it ignores the many important decisions practitioners must make. For example, any implementation of DP must at least specify the DP model (i.e., local, central, shuffle), the algorithm to be used (e.g., Gaussian noise or Laplace noise), and the values of $\epsilon$ and $\delta$. All of these decisions impact the practical privacy guarantees that data subjects receive, and should depend on the context and norms surrounding the application. Guidelines for context-specific appropriate choices of parameters is one of the largest open questions in DP practice today. The DP literature in computer science offers little contextual or normative guidance as to how these parameters should be tuned, while typically stating that they should depend on the context and the privacy-accuracy needs of the use-case (i.e., the contextual purposes). Our integrated approach will provide a formal framework for reasoning about the social norms and contextual purposes surrounding the choice of parameters in DP settings, and can provide guidance on appropriate context-dependent choices of these parameters.

\textbf{Continuous information design.} CI typically focuses on complete and exact information flows -- i.e., all specified information attributes about the subject flowing from the sender to the receiver, according to the transmission principle. Through the introduction of transmission properties, our integrated framework explicitly accounts for partial, aggregated, or noisy information flows. This includes, for example, partial information flows where the receiver only observes a noisy signal about the subject's information attributes, so any inferences made about the subject by the receiver will be partial, which is always the case for DP. The continuous information design within CI allows for a normative evaluation of the continuum of design choices to enable appropriate flows while limiting inappropriate flows.

\textbf{Aggregate information flows.} CI typically considers flows concerning a single data subject (e.g., a medical patient interacting with her healthcare provider). However, in many machine learning applications, the information contained in a flow pertains to many individuals (e.g., a prediction model trained on a large dataset of users). Extending CI to cover more real-world information flows enhances the richness and usability of the tool.

\textbf{Extending to other PETs.} While DP is our object of focus in this work, there are many other privacy-enhancing technologies (PETs) that could also benefit from contextual guidance on their deployments. Like DP, many common privacy and security tools from the computer science literature are algorithmic in nature without a contextual component; such examples include encrypted communication, secure multi-party computation, and federated learning. Our integrated approach also extends to other PETs beyond DP, to provide contextual guidance for the use and deployment of these other tools.

\section{Case study: Transmission Properties in the U.S. Census}

In this section, we analyze a Census-based case study using our integrated CI/DP framework. We find that CI and DP complement each other by clarifying
which contextual factors should be taken into account when determining appropriate choice of PETs and calibrating DP parameters. 

Specifically, we consider the release of data from the Decennial Census for the purposes of Congressional redistricting. Prior to 2020, the U.S. Census Bureau used a PET called \emph{swapping} to protect privacy of this data release; for the 2020 Decennial Census, the Census Bureau switched to using differential privacy instead. We analyze this case study through the lens of contextual integrity, and show that without articulating the \emph{transmission property} of this flow, it would be impossible to determine whether which PET leads to an appropriate information flow that meets the contextual purposes of the application respects the information norms.

We emphasize that the CI framework does not itself rule certain information flows appropriate or inappropriate, but rather creates a framework and language with which experts can determine appropriateness. As such, we do not aim to provide a ruling on which of these PETs is more appropriate, but rather show that the additional feature of transmission property is needed to highlight the context-relevant differences between these two PETs.

\subsection{U.S. Decennial Census}

The United States Decennial Census is a survey of the country's resident population that is conducted once every ten years. It is mandated by Section 2 of the U.S. Constitution \citep{USConst} in order to apportion seats to the House of Representatives. Since the first U.S. Decennial Census in 1790, the use of Census data products has expanded to be used for other purposes including redistricting, allocation of federal funds, and as a data resource for researchers.

Privacy requirements of U.S. Census Bureau data are derived from Title 13 of the United States Code, which imposes strict privacy requirements that must be met by the Bureau and its employees: they are sworn to protect the confidentiality of the data subjects of the census \citep{Title13}. Neither the Bureau nor any of its employees may ``make any publication whereby the data furnished by any particular establishment or individual under this title can be identified'' \citep{Title13}. The Bureau must meet its mandate of publishing meaningful statistics tabulated from the collected Decennial Census data, while still meeting the legal privacy requirements of Title 13.

Prior to the 2020 Decennial Census, the Census Bureau used a disclosure avoidance technique known as \emph{swapping} to preserve privacy, where it would swap individuals or their features in the database before publishing population counts by demographic and geographic breakdown. In 2020, the Bureau switched to using DP as its PET of choice to satisfy the privacy requirements of Title 13. This change was justified in part by work done by researchers at the Census Bureau showing that swapping did not sufficiently protect privacy, and that the majority of Americans could be re-identified from the published information even after swapping was applied \citep{Abo21,GAM18}.

There is an ongoing legal debate about whether DP or swapping is a more appropriate PET for the purposes of the U.S. Census \citep{nanayakkara2022s,BS22}. Viewed through the lens of contextual integrity, this debate is about the descriptive transmission properties of each PET, and whether they respect the established legal norms for protecting the privacy of individuals, and satisfy the contextual purposes of the Decennial Census. Thus without specifying the descriptive transmission properties under each PET as a part of the information flow, the framework of contextual integrity cannot be applied to evaluate whether the information flows with swapping or with DP respect contextual norms.

\subsection{The Context and Information Norms}\label{s.censuscontext}

Here we formalize the context and information norms associated with the Decennial Census. We recognize and acknowledge that the running the Decennial Census is a highly complex operation that produces many data products and involves a network of distinct information flows, many of which have different information norms. For this work, we consider a highly stylized model where the U.S. Census Bureau is a single entity (whereas in reality it comprised of many different individuals with different agent roles, ranging from enumerators to statisticians to security experts, and more), and we focus only on the publication of P.L. 94-171, which is the data product used for redistricting \citep{PL21}. We focus on this particular flow due to the interest in the Census Bureau's choice of PET in this application, and because of the existing legal norms surrounding privacy and usefulness of these data. We next formalize the CI parameters in this flow, and emphasize the role of a transmission property in determining appropriateness.

Recall that the existing definition of a context (Definition \ref{def.context}) includes: contextual purposes, agent roles, information attributes, information norms, and transmission principles. We show that using these parameters alone, the CI framework does not allow for a distinction between the information flows under swapping and DP. We then show that adding in notion of transmission properties does provide language that distinguishes between these two flows.

\textbf{Contextual Purposes. } Decennial Census data has its foundational purpose in the Constitution, to allocate seats in the House of Representatives to states based on population. In the modern era, this is done using the P.L. 94-171 Redistricting Data file \citep{PL21}. This is a result of Public Law 94-171, which ``requires the Census Bureau to provide states the opportunity to identify the small area geography for which they need data in order to conduct legislative redistricting'' \citep{PL21}. These geographic population counts are also used for redistricting at the state and local level, and for other federal allocation tasks. 

\textbf{Roles.} The agent roles in this stylized context are: the U.S. Census Bureau (sender), Persons residing in the U.S. (data subjects), Redistrictors (recipient), and General public (recipient).

\textbf{Attributes.} The information attributes about the data subjects included in the P.L. 94-171 are race, ethnicity, age, geographical location to the Census block level, and housing type (housing unit or group quarters).

\textbf{Transmission Principles.}
The normative transmission properties of this information flow are: it is compulsory for the data subjects (persons residing in the U.S.) to share their information attributes with the sender (U.S. Census Bureau) through the Decennial Census survey, and that it is required by law for these data files to be made publicly available.
Title 13 also specifies that individuals cannot be identifiable from the data products produced the by the U.S. Census Bureau.

\textbf{Information Norms.} The key information norm in this context is: the information attributes contained in the P.L. 94-171 files about all persons residing in the U.S. on the date of the Decennial Census may flow from the U.S. Census Bureau to the general public and redistrictors for the purpose of conducting population-based legislative redistricting, when the data were collected from the data subjects in the Decennial Census survey according to law, and the subjects cannot be re-identified from the release.

\subsection{The Role of Transmission Properties.} 

The ongoing legal debate around the use of DP in the Census \citep{nanayakkara2022s,BS22} is centered around the appropriateness of DP and swapping for these information flows. As specified in Section \ref{s.censuscontext} above, the context and the information norm associated with this data flow does not distinguish between the use of swapping and the use of DP, nor does it enable an answer to the question of which PET is appropriate for this application. It specifies the dual requirements that: (1) the data should enable accurate population-based redistricting, and (2) subjects cannot be re-identified. It does not include requirements on the algorithmic procedure for transforming the collected data into the released data -- as directly releasing the raw data without any PETs or disclosure avoidance methods would not be sufficient to prevent against re-identification attacks. It also does not specify technical requirements for meeting these goals; for example, what level of protection against re-identification attacks is sufficient to meet the privacy goal? These details must be reasoned about outside the framework of CI. Thus a CI analysis of the normative question of which PET is more appropriate for this information flow, could not give a complete answer without including the descriptive properties of the PET used to produce the data product. 

To fully address this question, we must include for consideration the two competing transmission properties of ``with swapping'' and ``with $(\epsilon,\delta)$-differential privacy''; these two information flows are indistinguishable under the language of the existing CI definition, but must be specified and treated differently in our augmented variant of information norms (Definition \ref{def.inormaug}). Recent research has shown that swapping cannot simultaneously satisfy the contextual purpose of accurate data for analysis and the information norm of non-identifiability of individuals \citep{Abo21, CRB22,BS22}, which suggests that swapping cannot be a part of an appropriate information norm in this context. It is widely agreed that DP satisfies the non-identifiability norm (for reasonably small values of $\epsilon$).  The ongoing debates about the use of DP for Census data products are centered around whether this PET also meets the contextual purpose of enabling accurate data analysis. An affirmative answer would provide a clear argument that DP can be used to produce appropriate information flows in this context that also meet the contextual purposes, whereas the previously employed method of swapping failed to do so.

We emphasize that the parameter-less transmission principle of ``with differential privacy'' without specifying values of $\epsilon$ and $\delta$ would also be insufficiently descriptive in this case, because these parameters determine the strength of the privacy guarantee. With $\epsilon=\infty$, no privacy protections are applied, and re-identification is very likely to be possible, thus violating the legal norm of preventing re-identification under Title 13. On the other hand, $\epsilon=0$ ensures perfect privacy, but the data release will not contain satisfactorily accurate information to facilitate the contextual purpose of redrawing legislative districts. Any transmission property of a DP information flow should specify the values of these parameters, as the flow may be appropriate under some $\epsilon$ values and inappropriate under others. This framework can then be used to map from the normative privacy requirements of a context to the descriptive mathematical parameterization of the privacy guarantees of a DP system, and thus provide provide normative guidance for the tuning of the $\epsilon$ parameter. Including the $\epsilon$ value in the description of the information flow can ensure that the chosen (algorithmic) transmission property satisfies the (normative) transmission principle's privacy requirements.

Beyond specifying the value of DP parameters $\epsilon$ and $\delta$, other parameters of the DP implementation are important to include in the transmission principle as well. These include the specific DP algorithm used (as many different algorithms can satisfy the same $(\epsilon,\delta)$ values while performing different operations on the data), the model of DP used (e.g., central, local, or shuffle), and whether the $(\epsilon,\delta)$ values are composed over multiple DP computations on the dataset. A complete specification of these algorithmic details should also be included in the transmission property of any information flow involving DP. We leave as future work the task of determining which differential privacy implementation details (including $\epsilon$ values) are appropriate in various contexts.

\section*{Acknowledgements}

The authors gratefully acknowledge feedback from reviewers, discussants, and participants of the venues where preliminary versions of this work appeared: Privacy Law Scholars Conference, the Theory and Practice of Differential Privacy Workshop, the Symposium on Applications of Contextual Integrity, and the USENIX Conference on Privacy Engineering Practice and Respect. We also acknowledge participants of the BIRS Workshop on Contextual Integrity for Differential Privacy -- especially Helen Nissenbaum -- for fruitful discussions related to the topic of this paper.

S.B. supported in part by NSF grants 2131532 and 2131532, and by the New York University Information Law Institute’s Fellows program, which is funded in part by Microsoft Corporation. R.C. supported in part by NSF grant CNS-1942772 (CAREER) and
an Early Career Faculty Impact Fellowship from Columbia University. Any opinions, findings, or conclusions expressed in this paper are those of the authors and do not necessarily reflect the views of the sponsors.

\bibliographystyle{plainnat}
\bibliography{privacy}

\end{document}